\documentclass[sigconf,10pt]{acmart}

\usepackage{xspace}
\usepackage{subfiles}
\usepackage{multirow}
\usepackage{tabularx}
\usepackage{subcaption}
\usepackage{enumitem}
\usepackage{balance}
\usepackage{xcolor}
\usepackage{cleveref}
\usepackage[splitrule,bottom]{footmisc}
\usepackage{longtable}
\usepackage{graphicx} 
\usepackage{rotating} 
\usepackage{lipsum}  
\usepackage{float} 
\usepackage{placeins} 
\usepackage{array} 
\usepackage{soul}
\usepackage{makecell}

\usepackage[normalem]{ulem}

\usepackage[font=small, skip=0.1ex]{caption}

\newcommand{\sysref}[1]{{\autoref{#1}}}

\AtBeginDocument{
  }

\setcopyright{none}
\copyrightyear{2025}
\acmYear{2025}
\acmDOI{XXXXXXX.XXXXXXX}

\acmConference[]{}{}

\acmISBN{978-1-4503-XXXX-X/2018/06}

\usepackage[subtle]{savetrees}
\begin{document}

\title[MoCoMR]{MoCoMR: A Collaborative MR Simulator with Individual Behavior Modeling}

\author{Diana Romero}
\email{dgromer1@uci.edu}
\affiliation{
  \institution{University of California, Irvine}
  \city{}
  \country{}}

\author{Fatima Anwar}
\email{fanwar@umass.edu}
\affiliation{
  \institution{University of Massachusetts, Amhrest}
  \city{}
  \country{}}

\author{Salma Elmalaki}
\email{salma.elmalaki@uci.edu}
\affiliation{
  \institution{University of California, Irvine}
  \city{}
  \country{}}

\newcommand{\sysname}{\textbf{MoCoMR}\xspace}

\renewcommand{\shortauthors}{Romero et al.}

\begin{abstract}
Studying collaborative behavior in Mixed Reality (MR) often requires extensive, challenging data collection. This paper introduces \sysname, a novel simulator designed to address this by generating synthetic yet realistic collaborative MR data. \sysname captures individual behavioral modalities such as speaking, gaze, and locomotion during a collaborative image-sorting task with 48 participants to identify distinct behavioral patterns.  \sysname simulates individual actions and interactions within a virtual space, enabling researchers to investigate the impact of individual behaviors on group dynamics and task performance. This simulator facilitates the development of more effective and human-centered MR applications by providing insights into user behavior and interaction patterns. The simulator's API allows for flexible configuration and data analysis, enabling researchers to explore various scenarios and generate valuable insights for optimizing collaborative MR experiences.
\end{abstract}



\maketitle

\section{Introduction}
Mixed Reality (MR) technologies are revolutionizing collaborative work by creating immersive environments that seamlessly blend digital and physical elements. These advancements offer unique opportunities to enhance teamwork, communication, and productivity across various domains. However, MR collaboration presents challenges in understanding group behavior due to the complex interplay of real and virtual elements, requiring new methods to capture and analyze these dynamics that are not fully addressed by existing research in immersive or traditional collaborative settings. 

Social graphs provide a structured approach to analyzing individual and group behavior in MR environments to model verbal exchanges, shared attention, and spatial relationships, offering deeper insights into group coordination and decision-making in MR environments~\cite{jia2024audio, romero2024groupbeamr}. However, current open-source datasets often fall short in meeting the requirements for studying both group dynamics and individual interactions in MR environments.

For instance, the Aria Synthetic Environments Dataset lacks human interactions, limiting its use for collaboration studies. Similarly, the Aria Digital Twin and Everyday Activities Datasets only cover single-user or dyadic interactions, missing the complexity of larger group settings~\cite{projectaria}. Other datasets, such as Stanford 2D-3D-Semantics and UCSD MR OpenRooms, primarily focus on static settings or limited interactions, failing to capture the intricacies of collaborative MR tasks~\cite{armeni2017joint,li2021openrooms}. Even the Egocentric Concurrent Conversations Dataset, which includes 50 participants, focuses solely on social conversations, excluding collaborative virtual tasks~\cite{ryan2023egocentric}. 

Although group behavior studies typically involve 3–6 members~\cite{krems2019conversations,lowry2006impact}, it is also crucial to examine individual dynamics within social settings. The inadequacy of existing datasets for studying both group- and individual-level behavior in MR environments motivates the creation of a simulation model that captures these complex dynamics, enabling more nuanced analyses of MR collaboration and social interaction.

To address the scarcity of tools for understanding human behavior in collaborative MR settings, we developed \sysname, a simulator designed to \uline{mo}del and analyze individual and group dynamics in \uline{co}llaborative \uline{MR} environments. \sysname is founded on a comprehensive framework for data collection, specifically tailored to capture the nuances of collaborative MR interactions. 

Our approach involved conducting an IRB-approved study with 48 participants, organized into 12 groups. Each group engaged in a collaborative task within a custom-designed MR environment, allowing us to capture a diverse range of behaviors and interactions. This rich dataset not only provides valuable insights into group dynamics in MR collaborative scenarios but also serves as the foundation for developing the simulation model of \sysname. By learning the underlying patterns and distributions of real-world collaborative behavior, \sysname enables the simulation of diverse scenarios and facilitates the exploration of various factors influencing individual and group performance in MR settings. 
\section{Related Work and Contribution}
Synthetic data enhances immersive technologies by improving perception models and tracking. In virtual reality, UnrealROX+ generates photorealistic datasets for tasks like semantic segmentation and object detection~\cite{martinez2021unrealrox+}. For augmented reality, synthetic techniques automate dataset creation for marker recognition~\cite{le2020machine}. EgoGen~\cite{Li_2024_CVPR} produces realistic first-person training data with simulated human motion for AR head-mounted displays. Synthetic data also supports head rotation prediction in XR~\cite{struye2022generating} and hand tracking in immersive applications~\cite{zhang2024vr}.

While previous works focus on synthetic data generation for visual perception and object tracking, they do not address the complexities of human behavior in collaborative MR environments. This work introduces a framework to understanding human behavior in collaborative MR by developing a simulator (\sysname) that simulates behavioral data. This contribution is significant because it addresses the challenge of obtaining and analyzing real-world collaborative MR data, which can be difficult and expensive to collect.

\sysname simulator focuses on eye gaze, audio, and location data, as these modalities offer a rich understanding of human behavior in shared virtual environments~\cite{romero2024groupbeamr}. Eye gaze shows attention, audio data captures communication and turn-taking, and location data tracks movement and spatial relationships. Modeling these modalities allows the simulator to provide APIs for simulating collaborative MR tasks and generating data that mirrors human interaction nuances, enhancing human-centered system design.This simulator can then be used to train and evaluate new adaptive systems and optimization algorithms, ultimately leading to improved human-centered design and implementation in immersive technologies. 
\section{Methodology}

\begin{figure*}
\centering
\includegraphics[width=\linewidth] {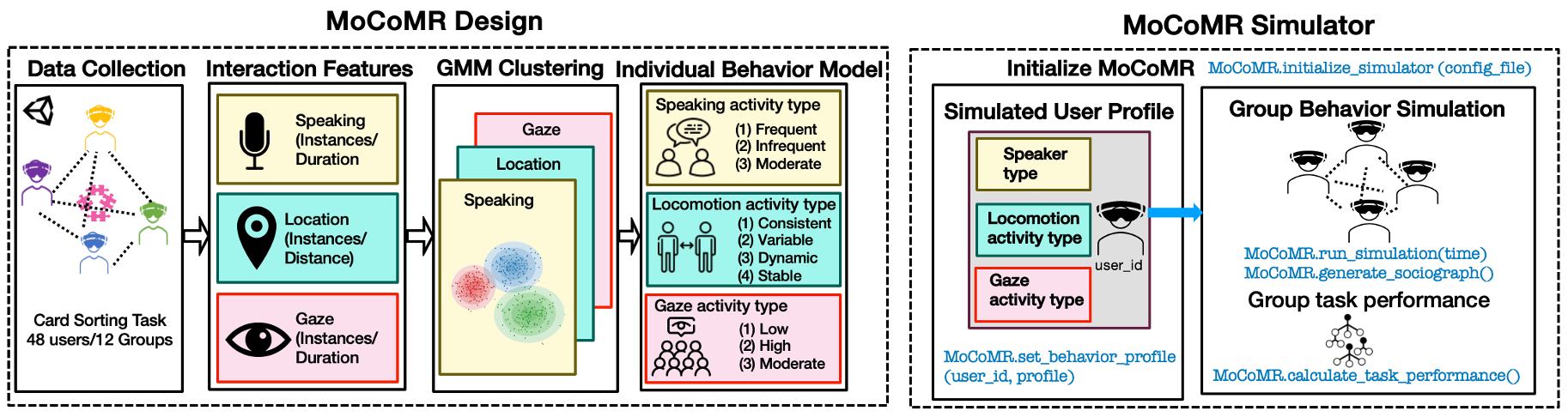}
\caption{Overview of the \sysname design pipeline. Features from various interaction types are used to identify emergent individual behaviors through Gaussian Mixture Model clustering. The identified behaviors serve as parameters to simulate individual types in our proposed \sysname simulator.} \label{fig:overview}
\end{figure*}

\sysname focuses on an image-sorting task~\cite{romero2024groupbeamr}, capturing key behavioral modalities such as location, gaze, speaking activity, and interactions with virtual objects. To achieve this, we employ a data-driven approach, learning the underlying distribution of our real-world dataset comprising 48 individuals engaged in a collaborative MR image-sorting task within 12 groups. By modeling individual behavior from this real-world data, we design \sysname simulator that can generate data that accurately reflects the statistical properties and behavioral patterns observed in actual collaborative MR settings. 
An overview of the \sysname design pipeline can be seen in \sysref{fig:overview}.

\subsection{Data Collection}

The study involved 48 participants (age 21-42, mean 24) in 12 groups of 4, using Meta Quest Pro headsets for a collaborative image sorting task. Participants categorized 28 OASIS dataset images into six emotion categories, designed to encourage decision-making, communication, and social coordination without time limits~\cite{kurdi2017introducing,yangImmersiveCollaborativeSensemaking2022}. Measurements included sensor-level data (audio, location, eye-tracking), task-level metrics (virtual object interactions, label changes), and performance metrics (completion time, subjective accuracy). The IRB-approved study, with participants' informed oral consent, aimed to observe open collaboration in MR, focusing on interaction, task performance, and subjective experiences to analyze group behavior and validate the proposed system.

\subsection{Individual Behavior Modeling}
Our approach to modeling individual behaviors in MR environments employs Gaussian Mixture Model (GMM) clustering for each interaction type: speaking, gaze, and location. This method allows for nuanced representation and analysis of human behavior in collaborative MR tasks.

The clustering process involves three key steps:

\begin{enumerate}[leftmargin=0.6cm, topsep=0.1cm]

    \item[\textbf{(a)}]  \emph{Data Preparation:} Raw interaction logs are parsed to extract time-series data for each interaction type (speaking, gaze, and location). This includes details like speaking instances, gaze fixations, and headset movement data.
    
    \item[\textbf{(b)}]  \emph{Feature Extraction:} Statistical and frequency-domain features are derived to capture individual interaction characteristics. This includes temporal statistics and Fourier Transform to extract frequency-based patterns. \sysref{tab:features} summarizes the features used.
    
    \item[\textbf{(c)}]  \emph{Model Fitting:} Multiple Gaussian Mixture Models (GMMs) are evaluated using Bayesian Information Criterion (BIC), Silhouette Score, and Davies-Bouldin Index (DBI) to determine optimal cluster numbers. The final model is selected based on the knee-point of the BIC curve. The resulting clusters represent distinct behavioral patterns in speaking frequency, gaze fixation tendencies, and movement characteristics, which form the basis for simulating individual behaviors in the collaborative MR environment.

\end{enumerate}

\begin{table}
\centering
\scriptsize
\caption{Features used for clustering each interaction type}
\label{tab:features}
\begin{tabular}{|l|p{6cm}|}
\hline
\textbf{Interaction Types} & \textbf{Features}  \\ \hline \hline
    
\textbf{Speaking} & 
Instances count, mean/std of timestamps and duration, first three non-DC Fourier frequency components of timestamps. \\ \hline

\textbf{Gaze} & 
Instances count, mean/std of timestamps and duration, virtual object looked at, first three non-DC Fourier frequency components of timestamps. \\ \hline

\textbf{Location} & 
Instances count, mean/std of timestamps and X, Y, Z coordinates, range of X, Y, Z, total distance/time, mean/max speed and acceleration, mean jerk, path tortuosity, idle fraction, first three non-DC Fourier frequency components. \\ \hline

\end{tabular}
\end{table}

\subsection{Individual Behavior Simulation and Group Interaction Modeling}

\sysname uses GMM cluster characteristics to model and simulate different interaction behaviors of a user. The process involves recreating speaking, gaze, and location behaviors while maintaining consistency with observed patterns in collaborative MR setup creating simulated behavioral sequences statistically similar to real-world data.

\sysname takes cluster configurations for each interaction type as input, dictating simulated participant behavior. It outputs log files replicating location movement, speaking patterns, gaze behavior, and virtual object interactions for each participant. The simulation includes:

\begin{enumerate}[leftmargin=0.6cm, topsep=0.1cm]
    \item[\textbf{(a)}]  \emph{Locomotion  Simulation:}  Movement data is generated by \sysname by sampling locomotion patterns from clustered participant trajectories. A multivariate Gaussian model is fitted to spatial and temporal transitions, enabling the synthesis of realistic movement sequences. Timestamp adjustments ensure temporal coherence, and synthetic location logs are formatted to match the original headset position recordings.

    \item[\textbf{(b)}]  \emph{Speaking Simulation:} Speaking behaviors are recreated using frequency-based sampling techniques. Speaking instances are generated based on Fourier-transformed start time distributions and duration histograms. \sysname generates speaking logs by sampling from smoothed distributions of observed speaking patterns, preserving the rhythmic characteristics of natural conversations.

    \item[\textbf{(c)}]  \emph{Gaze Simulation:} Eye-tracking behaviors are simulated using captured gaze data distributions. Virtual object fixations are determined through reconstructed frequency-domain features from the original gaze data. \sysname generates synthetic gaze logs by applying inverse Fourier transforms to approximate gaze event distributions, ensuring the timing and duration of gaze behaviors reflect real-world tendencies.
\end{enumerate}

\sysname APIs is used to simulate a four-member group image sorting task. The interaction data generated is used to create structured representations of group dynamics, including directed conversation graphs for turn-taking and speaking dominance, undirected proximity graphs for spatial interactions, and shared attention graphs for gaze synchrony.

We utilize a machine learning framework to examine how individual behaviors relate to group virtual object interactions and task metrics. This approach aggregates the speaking, location, and gaze clusters for each participant, computing the most representative cluster per group. We then link this with object interaction and task performance metrics from observational data.

In particular, a \texttt{RandomForestRegressor} model is trained to predict task metrics from cluster configurations using one-hot encoded categorical cluster IDs. Its predictive performance is assessed by Mean Absolute Error (MAE) and R-Squared ($R^2$) scores to evaluate the gap between real-world and simulated data.

Through this process of simulating individual behaviors and synthesizing group interactions, we generate structured social interaction data that aligns with observed patterns.

\subsection{\sysname API Design}

The \sysname simulator offers a user-friendly API for conducting and analyzing simulations.

\begin{itemize}[noitemsep, leftmargin=*, topsep=0pt]
\item \textbf{Initialization:}  \texttt{initialize\_simulator(config\_file)} sets up the simulation environment, including virtual users, environment settings, and task parameters.

\item \textbf{Behavior \& Task Setup:} \texttt{set\_behavior\_profile(user\_id, profile)} defines user behaviors, while \texttt{set\_task(task\_type, parameters)} configures the collaborative task.

\item \textbf{Simulation Execution:} \texttt{run\_simulation(duration)} executes the simulation. 

\item \textbf{Data Retrieval:}  \texttt{get\_user\_data(user\_id)} provides individual data and \texttt{get\_group\_data()} provides group-level data.

\item \textbf{Analysis \& Output:} \texttt{generate\_sociograph()} creates a sociograph, \texttt{calculate\_task\_performance()} computes performance metrics, and \texttt{export\_data(format)} exports data for further analysis.
\end{itemize}

This API provides researchers with a powerful toolset for conducting controlled experiments, analyzing user behavior, and evaluating different design choices in collaborative MR environments.

\section{Evaluation}

To validate our simulation, we use \sysname to create data for 48 users in 12 groups of 4, reflecting observational data. Each simulated user follows GMM cluster assignments derived from real-world interactions. Individual behaviors are assessed through instances, durations, histogram similarity, Wasserstein distance, and correlation. Group interactions are evaluated using graph metrics like Jaccard and cosine similarity, mean weight, interaction differences, and graph isomorphism. These metrics determine how accurately the simulation reflects observed collaboration in MR settings.

\subsection{Individual Speaking Behaviors}

\begin{figure}
\centering
\includegraphics[width=0.7\linewidth]{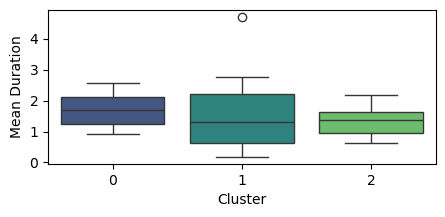}
\caption{Distribution of mean speaking duration by cluster.} \label{fig:speaking-cluster}
\end{figure}

The GMM clustering identified three optimal clusters based on the Bayesian Information Criterion (BIC). The silhouette score of $0.21$ indicates weak separation between clusters, while the Davies-Bouldin Index (DBI) of $1.44$ suggests moderate clustering quality with some overlap between groups. The clusters of individual speaking data can be categorized into the following characteristics, as illustrated in \sysref{fig:speaking-cluster}:

\begin{enumerate}[leftmargin=0.6cm, topsep=0.1cm]
    \item[\textbf{(a)}]  \emph{Frequent Talkers (Cluster 0):} Exhibited the highest mean number of speaking instances (119.29) with moderate speaking duration (1.7 seconds).

    \item[\textbf{(b)}]  \emph{Infrequent Talkers (Cluster 1):} Demonstrated the lowest mean number of speaking instances (14.35) with consistent but short speaking durations (mean = 1.6 seconds).

    \item[\textbf{(c)}]  \emph{Moderate Talkers (Cluster 2):} Showed an intermediate mean number of speaking instances (48.63) with the shortest mean speaking duration (1.32 seconds) and moderate variability.
\end{enumerate}

\begin{figure}
\centering
\includegraphics[width=\linewidth]{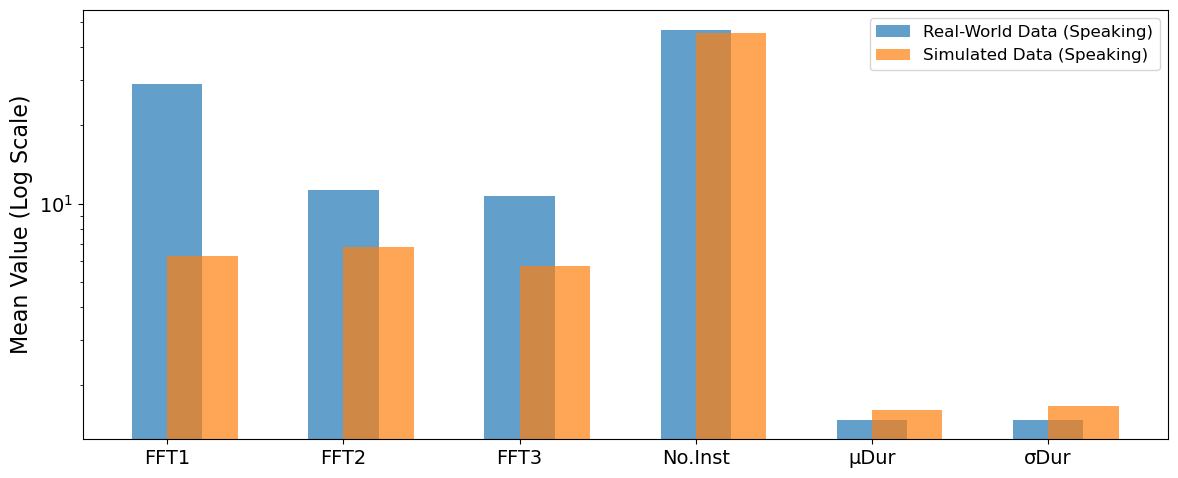}
\caption{Real-world and simulated mean speaking features.} \label{fig:speaking-eval}
\end{figure}

We evaluated model fidelity by comparing real-world and simulated speaking behaviors \sysref{fig:speaking-eval}. The simulated speaking instances (mean: 45.35) closely match the real data (mean: 46.79), with high histogram similarity (0.83) and low Wasserstein score (0.23). Autocorrelation score (0.85) indicates maintained temporal patterns. Mean durations in simulations (1.60) align closely with real data (1.47), with good similarity (0.69) and low Wasserstein (0.17). Deviations appear in Fourier transform features (FFT1, FFT2, FFT3), mainly in FFT1, but histogram similarities (0.87 to 0.92) and low Wasserstein scores (0.08 to 0.13) indicate preserved periodic patterns, despite lower magnitudes.

Overall, the simulated data demonstrates strong alignment with real-world data across most features, with slight deviations in frequency components. The consistent use of histogram similarity and Wasserstein distance metrics confirms that our proposed modeling in \sysname is able to generate the simulated data effectively to approximate the real-world speaking behaviors.

\subsection{Individual Gaze Behaviors}

\begin{figure}
\centering
\includegraphics[width=0.7\linewidth]{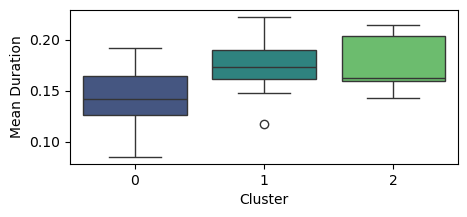}
\caption{Distribution of mean gaze duration by cluster.} \label{fig:gaze-cluster}
\end{figure}

The GMM clustering of participant gaze behavior data identified three clusters, with the optimal number determined by BIC and KneeLocation method. The silhouette score of $0.29$ and DBI of $1.13$ suggest that the clusters are somewhat distinct but not perfectly separated. \autoref{fig:gaze-cluster} illustrates the characteristics of each cluster:

\begin{enumerate}[leftmargin=0.6cm, topsep=0.1cm]
    \item[\textbf{(a)}]  \emph{Low Gaze Activity (Cluster 0):} Lowest number of gaze instances (mean = 836.43), and shortest mean duration (0.14) with lowest standard deviation (0.18).

    \item[\textbf{(b)}]  \emph{Moderate  Gaze Activity (Cluster 1):} Moderate gaze activity (mean = 1325.77 instances), and slightly higher mean duration (0.18) and standard deviation (0.24) than Cluster 0.

    \item[\textbf{(c)}]  \emph{High Gaze Activity (Cluster 2):} Highest number of gaze instances (mean = 2014.00), with a mean duration similar to Cluster 1 (0.18), but highest standard deviation (0.25).
\end{enumerate}

\begin{figure}
\centering
\includegraphics[width=\linewidth]{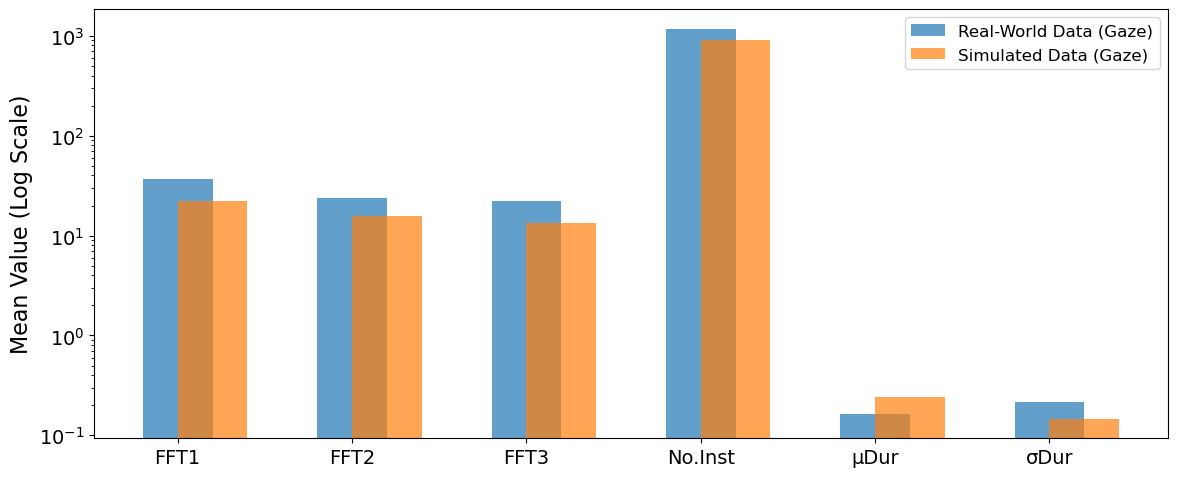}
\caption{Real-world and simulated mean gaze features.} \label{fig:gaze-eval}
\end{figure}

We compared simulated and real-world gaze behaviors using key features in \sysref{fig:gaze-eval} on a log scale. Simulated data exhibits real-world behaviors but with fewer instances (mean = 907.5 vs. 1183.38) and differing distributions (histogram similarity = 0.31, Wasserstein similarity = 0.0034). Autocorrelation (0.67) and partial autocorrelation (0.77) moderately maintain temporal patterns. Simulated data closely replicates mean duration distribution (simulated mean = 0.24, real-world mean = 0.16) with high histogram (0.083) and Wasserstein (0.958) similarities. Although FFT features are lower, high autocorrelation scores (0.92 to 0.95) indicate preserved periodicity.

Overall, the simulated data demonstrates moderate to high similarity with the real-world data, particularly in terms of duration, variability, and periodicity, despite discrepancies in the number of instances. These results suggest that the simulated data effectively capture the main dynamics of real-world speaking behaviors, providing a reliable approximation for simulation purposes.

\subsection{Individual Locomotion Behaviors}

\begin{figure}
\centering
\includegraphics[width=0.7\linewidth]{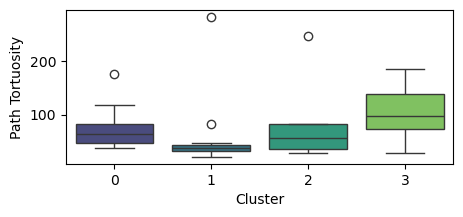}
\caption{Distribution of path tortuosity by cluster.} \label{fig:gaze-cluster}
\end{figure}

The GMM clustering of the locomotion behavior data identified four distinct clusters, with the optimal number determined by the BIC and the KneeLocator method. The silhouette score of 0.11 and Davies-Bouldin Index (DBI) of 1.79 suggest that while the clusters are not perfectly separated, they capture meaningful distinctions in movement dynamics, including variations in path tortuosity, speed, acceleration, and idle fraction. The differences in path tortuosity across the clusters are shown through boxplots in \sysref{fig:gaze-cluster}, which illustrate the varying levels of movement complexity within each cluster.

\begin{enumerate}[leftmargin=0.6cm, topsep=0.1cm]
    \item[\textbf{(a)}]  \emph{Moderate Consistent Locomotion (Cluster 0):} Lowest number of instances (mean = 496.17), moderate path tortuosity (mean = 2.49), stable mean speed (max 497.34), low jerk and acceleration values, and low idle fraction (0.16). Movement is predictable and consistent.

    \item[\textbf{(b)}]  \emph{Higher Variable Locomotion (Cluster 1):} Moderate path tortuosity (mean = 3.45), higher max speed (859.66), moderate jerk and acceleration values, and slightly higher idle fraction (0.18). More winding and erratic movement with frequent changes in speed and direction.

    \item[\textbf{(c)}]  \emph{Highest Dynamic Locomotion (Cluster 2):} Largest number of movement instances (mean = 1140.2), highest path tortuosity (mean = 2.29), highest max speed (1146.38), very high jerk and acceleration values, and low idle fraction (0.14). Quite complex and winding paths with rapid changes in pace and direction.

    \item[\textbf{(d)}]  \emph{Moderate Stable Locomotion (Cluster 3):} Intermediate level of movement (mean instances = 709.18), moderate path tortuosity (mean = 2.32), moderate max speed (710.63), and low idle fraction (0.14). More predictable movement with steady, less erratic patterns.
\end{enumerate}

FFT features distinguish the clusters: Cluster 0 has the lowest values (FFT1 = 31.13, FFT2 = 19.47, FFT3 = 16.71), indicating low periodicity, while Cluster 2 has the highest (FFT1 = 49.82, FFT2 = 42.64, FFT3 = 38.29), suggesting more regular movement patterns. Clusters 1 and 3 have intermediate values, indicating moderate periodicity. These results show that the GMM clustering method effectively captures diverse locomotion behaviors, offering insights into how individuals interact with their environment through movement.

\begin{figure}
\centering
\includegraphics[width=\linewidth]{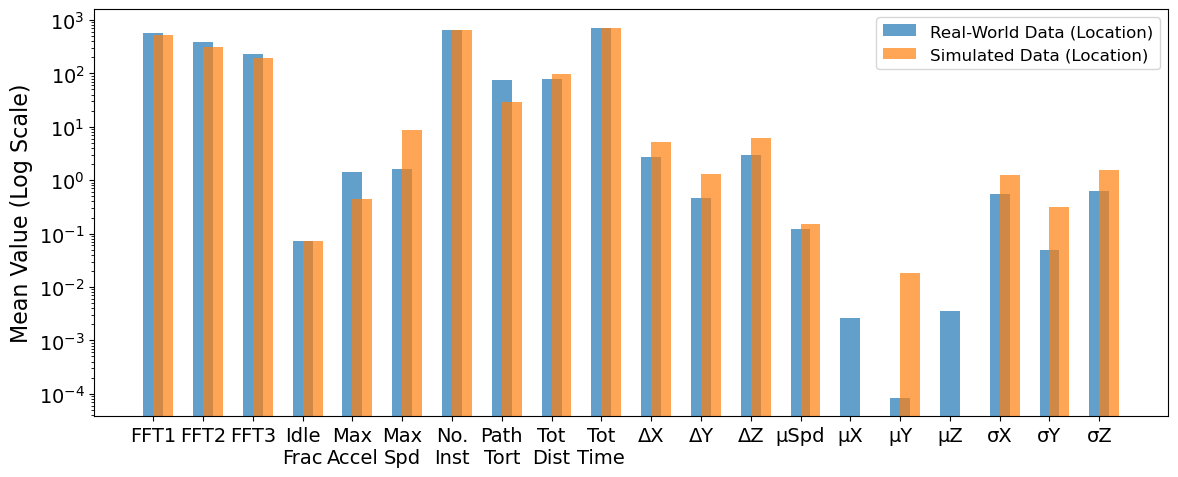} 
\caption{Real-world and simulated mean location features.} \label{fig:location-eval}
\end{figure}

We analyzed real and simulated locomotion behaviors in log scale per \sysref{fig:location-eval}. The simulation aligns with real data in some areas but has discrepancies. The simulated mean (647.77) is close to the real mean (648.13), with histogram and Wasserstein similarities (0.29 and 0.018) indicating a similar distribution. Autocorrelation (0.87) and partial autocorrelation (0.89) suggest the temporal structure is mostly preserved. Simulated spatial coordinates (X, Y, Z) show higher variability, with moderate histogram similarity for X and Y (0.31 to 0.51). Simulated paths are less tortuous (28.78) than real ones (76.47), showing simpler movement. FFT analysis reveals lower periodicity compared to real data. Despite this, overall correlation (0.74) and autocorrelation scores (0.86 to 0.92) indicate that simulated data generally mimics real-world characteristics, despite differences in complexity and periodicity.

In summary, while the simulated data captures some key aspects of the real-world location behaviors, it shows limitations in replicating the full complexity of movement patterns and spatial distributions. These findings highlight areas for potential improvement for \sysname in modeling the locomotion behavior in collaborative MR.

\subsection{Individual Object Interaction Behaviors}

\begin{table}[!t]
\caption{Summary of average errors between real-world and simulated task metrics. }
\scriptsize
\centering
\begin{tabular}{|c|c|}
\hline
{\textbf{Metric}} & \textbf{Average \% Error} \\ \hline
{No. of Images Grabbed \% Error} & 8.47 \\ \hline
{Total No. of Image Grabbing \% Error} & 14.21 \\ \hline
{No. of Image Labels Overridden \% Error} & 28.60 \\ \hline
{Total Images Looked At \% Error} & 0.63 \\ \hline
{Completion Time (seconds) \% Error} & 18.09 \\ \hline
{Accuracy (\%)} \% Error & 24.64 \\ \hline
{No. of Label Changes \% Error} & 12.52 \\ \hline
\end{tabular}
\label{tab:percent_errors_transposed_rounded}
\end{table}

The RandomForestRegressor model was used to predict task performance metrics based on group-level virtual object interaction behaviors, with task metrics such as accuracy, completion time, and the number of interactions. The model's Mean Absolute Error (MAE) was 39.15, indicating that, on average, the model's predictions were off by this value. The R-Squared score of -0.4348 suggests that the model did not effectively capture the relationship between group interaction patterns and task outcomes, as the negative value indicates poor explanatory power.

Table \ref{tab:percent_errors_transposed_rounded} shows varying prediction accuracy levels for task metrics. Completion time and images viewed had low errors ($18.09\%$ and $0.63\%$), while accuracy and label changes had higher errors, reflecting prediction challenges. Overall, the results highlight the difficulty of accurately predicting task outcomes based on group interaction behaviors in \sysname modeling approach.

\subsection{\sysname Simulated Data Evaluation}

\begin{table}[!t]
\centering
\scriptsize
\caption{Simulated data evaluation results. }
\label{tab:simulation_results}
\begin{tabular}{|p{3.5cm}|p{1cm}|p{1cm}|p{1.75cm}|}
\hline
\textbf{Metric} & \textbf{Conv.} & \textbf{Prox.} & \textbf{Shared Att.} \\
\hline
Mean Weight Diff & 0.20 & 0.56 & 0.31 \\ \hline
Mean Node Interaction Diff & 0.57 & 0.86 & 0.55 \\ \hline
Jaccard Edges & 0.98 & 0.86 & 0.96 \\ \hline
Cosine Similarity & 0.91 & 0.64 & 0.93 \\ \hline
Graph Isomorphism & 0.92 & 0.58 & 0.92 \\ \hline
\end{tabular}
\end{table}

We analyzed group-level interactions using graph-based metrics. We simulated four-member group interactions based on GMM cluster assignments from real-world data, focusing on conversation, proximity, and shared attention dynamics. These were represented as sociograms: verbal interaction (speaking duration), gaze-based (shared attention on virtual objects), and proximity-based (spatial relationships within 1.5 feet). \sysref{tab:simulation_results} summarizes the comparison between observed and simulated data sociograms.

The results demonstrated strong alignment between simulated and observed behaviors. Conversation graphs showed high fidelity, with an average Jaccard similarity of 0.93, high cosine similarity (mean = 0.83), and graph isomorphism (mean = 0.81) scores. All 12 groups maintained perfect Jaccard edge similarity in most cases, with minimal variations in mean node interaction and relative weight differences.

\begin{figure}[t]
  \centering
  \begin{subfigure}[b]{0.3\linewidth}
    \centering
    \includegraphics[width=\linewidth]{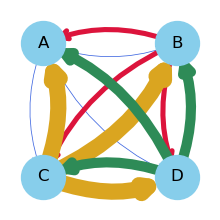}
    \caption{}  
    \label{fig:conv-actual}
  \end{subfigure}
  \begin{subfigure}[b]{0.3\linewidth}
    \centering
    \includegraphics[width=\linewidth]{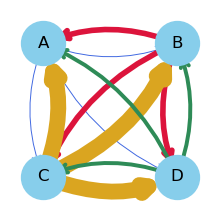}
    \caption{ }  
    \label{fig:conv-synth}
  \end{subfigure}
  \caption{Sample conversation sociogram for (a) actual observed data and (b) \sysname simulated data with the same participant configuration.}
  \label{fig:socio-sample}
\end{figure}

Proximity interactions showed moderate to strong alignment, with an average Jaccard similarity of 0.73, though some groups were lower (0.5-0.6). Proximity graph isomorphism was generally high (mean = 0.8), indicating strong spatial replication. Shared attention graphs had a Jaccard similarity of 0.93, a high cosine similarity (mean = 0.83), and consistently high graph isomorphism, except for group 10. Graphs in \sysref{fig:socio-sample} display real-world and \sysname simulated speaking behaviors. 

The simulation by \sysname accurately captured group interaction dynamics' structure and timing, with minor spatial and attention discrepancies. Using \sysname, this approach models group dynamics in MR environments, providing valuable insights into collaborative behaviors.
\section{Discussion \& Conclusion}
This work introduces \sysname, a simulator for individual and group behavior in collaborative MR environments. Evaluation of \sysname demonstrated its ability to capture structural and temporal aspects of group interaction dynamics, with minor discrepancies in spatial proximity and shared attention patterns. The validation highlights the simulator's ability to provide insights into collaborative MR experiences and guide the creation of effective human-centered applications.

Varying prediction accuracies highlight the challenges of modeling complex human behaviors. Metrics such as completion time and images viewed had low errors, whereas accuracy and label changes had higher errors due to subjectivity. Future work will focus on investigating correlations between different interaction types and collaborative task to enhance the fidelity of simulated data. By understanding how various behavioral modalities (e.g., speaking, gaze, movement) influence each other, researchers can create more nuanced and realistic simulations.

The significance of this work is particularly evident in the context of data-hungry models in MR and human-computer interaction research. By providing a means to generate large amounts of simulated data that closely mimics real-world interactions, \sysname addresses a critical need in the research community. This approach can accelerate the development and testing of new algorithms, interaction designs, and adaptive systems for collaborative MR environments.

As MR technologies continue to evolve and become more prevalent across various domains, tools like \sysname will play a crucial role in shaping the future of human-centered design in immersive collaborative environments.

\begin{acks}
This work is supported by the U.S. National Science Foundation (NSF) under grant number 2339266.
\end{acks}

\bibliographystyle{ACM-Reference-Format}

\bibliography{paper}

\end{document}